\documentstyle[twocolumn,aps,epsf]{revtex}

\begin{document}

\hyphenation{presence repeatedly experiments several YBCO LSCO sintered NMR}

\twocolumn[
\hsize\textwidth\columnwidth\hsize\csname@twocolumnfalse\endcsname
\draft

\title{Charge Localization from Local Destruction of Antiferromagnetic
Correlation in Zn-doped YBa$_{2}$Cu$_{3}$O$_{7-\delta}$}
\author{Kouji Segawa and Yoichi Ando}
\address{Central Research Institute of Electric Power
Industry, Komae, Tokyo 201-8511, Japan}
\date{Received 29 July 1998}
\maketitle

\begin{abstract}
The in-plane normal-state resistivity of Zn-doped
YBa$_{2}$Cu$_{3}$O$_{7-\delta}$ single crystals
is measured down to low temperatures by suppressing superconductivity
with magnetic fields up to 18 T.
Substitution of Cu with Zn in the CuO$_2$ planes is found to
induce carrier localization at low temperatures in \lq\lq clean"
samples with  $k_{F}l > 5$, where the mean free path $l$ is
larger than the electron wave length and thus localization
is not normally expected.
The destruction of the local antiferromagnetic correlation
among Cu spins by Zn is discussed to be the possible origin
of this unusual charge localization.
\end{abstract}

\pacs{PACS numbers: 74.25.Fy, 74.62.Dh, 74.20.Mn, 74.72.Bk}

]

After more than 10 years of intense research, the mechanism of
high-$T_{\rm c}$ superconductivity remains to be elucidated,
as well as the origin of the peculiar normal-state properties
of the high-$T_{\rm c}$ cuprates \cite{Maple}.
It has been repeatedly pointed out that the antiferromagnetic (AF)
spin correlation in the CuO$_2$ planes may play a
fundamental role in the normal-state properties and the
superconductivity.
Recently, there appeared several experiments which
show AF correlation is likely to be fundamentally relevant to the
high-$T_{\rm c}$ superconductivity.
For example, incommensurate spatial modulation of AF fluctuation
has been reported in superconducting La$_{2-x}$Sr$_{x}$CuO$_{4}$ (LSCO)
\cite{Yamada} and in YBa$_{2}$Cu$_{3}$O$_{7-\delta}$ (YBCO)
\cite{Mook},
and the incommensurability is found to be closely related to
$T_{\rm c}$ \cite{Yamada}.

The AF correlation in the CuO$_2$ planes can be disturbed
by partially replacing Cu with other atoms.
It is known that partial substitution of Cu (which has spin 1/2)
with Zn (which is nonmagnetic) destroys
local AF correlation and thereby quite drastically
suppresses superconductivity \cite{Mahajan}.
On the other hand, the effect of Zn substitution on the normal-state
charge transport has been believed to be rather modest;
it is reported that Zn-substitution simply adds some residual scattering
\cite{Chien,Fukuzumi} and leaves the signature of the
pseudogap in transport properties unchanged \cite{Mizuhashi}.
Therefore, it would be interesting to look for some fundamental change
in the normal-state charge transport caused by the destruction
of the local AF correlation upon Zn-doping.
This becomes particularly intriguing in the light of the recently
observed logarithmic divergence of the normal-state resistivity
in La$_{2-x}$Sr$_{x}$CuO$_{4}$ (LSCO) \cite{Ando},
in which system presence of a dynamical charge stripe order is discussed
\cite{Tranquada,Suzuki}.
The charge stripes intervene in the antiferromagnetically correlated
domains, so that
the stripes are one-dimensional (1D) defects to the AF correlation;
similarly, Zn is a point-like defect to the AF correlation.
If the charge stripe in the LSCO system is the source of the unusual
increase in resistivity at low temperatures as recently proposed
\cite{Tranquada2}, then Zn-doping might also induce similar
unusual localization behavior at low temperatures, although the
dimensionality of the defect is different.

In this paper, we report measurement of the low-temperature
normal-state resistivity along the CuO$_2$ planes, $\rho_{ab}$,
of Zn-doped YBCO single crystals by suppressing superconductivity
with dc magnetic fields up to 18 T.
We found that the Zn substitution induces unusual carrier
localization at low temperatures in a \lq\lq clean"
system with $k_{F}l > 5$, where $l$ is the mean free path and
$k_F$ is the Fermi wave number.
This observation strongly suggests that
the local destruction of the AF correlation
by Zn-doping can severely and qualitatively affect the charge transport
in cuprates; therefore, the AF correlation in the
CuO$_2$ plane seems to be indispensable to the peculiar metallic
normal-state charge transport.

There are two different Cu sites in YBCO, the chain site Cu(1)
and the plane site Cu(2); when Zn is doped to YBCO, it is reported that
Zn replaces mostly Cu(2) atoms, leaving the carrier density unchanged
\cite{Alloul}.
NMR measurements have revealed that nonmagnetic Zn induces local
magnetic moment by causing a spin \lq\lq hole" in the AF background of the
Cu spins \cite{Mahajan}.  This local moment gives rise to
a Curie term in the magnetic susceptibility at low temperatures
\cite{Alloul}.
In recent years, the effects of Zn-doping on the
pseudogap phenomena in the underdoped
YBCO have been extensively investigated.
Essentially, physical probes that are sensitive to the excitations with
${\bf q}$=$(\pi,\pi)$ [$(T_1 T)^{-1}$ in NMR,
inelastic neutron scattering, etc.]
find that Zn-substitution diminishes the pseudogap, while the probes
that are only sensitive to excitations with ${\bf q}$=$(0,0)$ (magnetic
susceptibility, NMR Knight shift, electrical resistivity, etc.) find that
pseudogap signatures are intact upon Zn-substitution \cite{Mizuhashi}.
The two-dimensional (2D) superconductor-insulator (S-I) transition,
which occurs at a critical sheet resistance (per CuO$_2$ plane) of
$h/4e^2$ (=6.5 k$\Omega$), has also been studied in Zn-doped YBCO
\cite{Fukuzumi,Walker}.
In this transition, Cooper pairs are conjectured to localize in the
presence of disorder \cite{Fisher}; Zn in this case is
considered to introduce disorder potentials in the usual sense,
though the scattering from Zn is maximally large (unitarity limit)
\cite{Mizuhashi,Nagaosa}.
Specifically, the S-I transition in YBCO takes place at around
400 $\mu \Omega$cm in single crystals \cite{Fukuzumi}
(a larger critical value of 750 $\mu \Omega$cm is reported for thin films
\cite{Walker}).

The single crystals of YBa$_{2}$(Cu$_{1-z}$Zn$_z$)$_3$O$_{7-\delta}$
are grown by flux method in pure
Y$_2$O$_3$ crucibles to avoid inclusion of any impurities other than Zn.
The purity of the crucible and the starting powders are 99.9 \% and
99.95 \%, respectively.
All the crystals measured here are naturally twinned.
The oxygen content $y$ ($\equiv$ 7-$\delta$) in the crystals is
controlled by annealing in evacuated and sealed quartz tubes
at 500 - 600$^{\circ}$C for 1 - 2 days
together with sintered blocks and powders,
for which the oxygen content is controlled
beforehand.
The final oxygen content is confirmed by
iodine titration method and also by measuring
the weight of the sintered blocks.
The obtained values from the two techniques are consistent
within $\pm 0.02$.
To obtain sharp superconducting transition,
the quartz tubes are quenched with liquid nitrogen at the end of
the annealing.  Although it has been discussed that quenching from a high
temperature causes oxygen disorder in the CuO chains \cite{Mizuhashi},
the residual resistivities of the quenched samples and slowly-cooled
samples are identical for our crystals.
Because of the careful control of the purity and the oxygen content,
the variation of $\rho_{ab}$ and $T_c$ is less than 5\% and 2 K,
respectively, and our data are in good agreement
with the best data in the literature \cite{Fukuzumi,Mizuhashi}.
The measurements of $\rho_{\rm ab}$ are performed
with ac four-probe technique
under dc magnetic fields up to 18 T applied along the $c$ axis.

The actual concentration of Zn in the
YBa$_{2}$-(Cu$_{1-z}$Zn$_z$)$_3$O$_y$ crystals is
measured with inductively-coupled plasma spectrometry (ICP)
with an error in $z$ of less than $\pm$0.10\%.
To supplement the ICP result, we compared $T_c$ of
fully oxygenated crystals with that of sintered samples
where Zn concentrations are known.
The results from the two techniques are consistent with each other.
The concentration of Zn in the crystals thus determined is roughly the
half of the content before the crystal growth.
The homogeneity of Zn in the crystals is confirmed with
electron-probe microanalysis (EPMA).

Figure 1 shows the temperature dependence of $\rho_{\rm ab}$ in
0 and 18 T, for three samples with different Zn concentrations.
Note that the three samples have the same oxygen content of 6.70.
In zero field, all the samples are metallic ($d\rho/dT$$\ge$0)
and show superconductivity.
We can see in Fig. 1 that the effect of
Zn-doping in zero field is to reduce $T_c$ and simply add
some constant residual resistivity (thereby shifting up the
$\rho_{ab}(T)$ curve),
as is already reported in the literature.
Thus, there is no qualitative difference among the three samples in the
zero-field properties.
However, under a high magnetic field of 18 T, a drastic
difference emerges; $z$=2.7\% sample shows a clear upturn in
$\rho_{\rm ab}$ at low temperatures, while other samples with smaller
$z$ do not show such clear upturn.
We also prepared a slowly-cooled sample to check if the
observed localization behavior ($d\rho/dT$$<$0) in the $z$=2.7\%
sample is due to oxygen disorder; essentially the same localization
behavior is observed in the slowly-cooled sample, which is supposed to
have smaller amount of oxygen disorder (at the expense of transition
width).
Thus, for the oxygen content of 6.70, we may conclude that
2.7\% of Zn induces carrier localization in YBCO
at low temperatures in 18 T,
irrespective of the annealing procedure.
The inset to Fig. 1 shows a plot of $\rho_{ab}$ vs $\log T$
for the 18-T data of $z$=2.7\% sample;
the temperature dependence of $\rho_{ab}$ is consistent with
$\log(1/T)$, as in the case of the underdoped LSCO \cite{Ando}.

\vspace{-0.7cm}
\begin{figure}[htbp]
\begin{center}
 \epsfxsize=75mm
 $$\epsffile{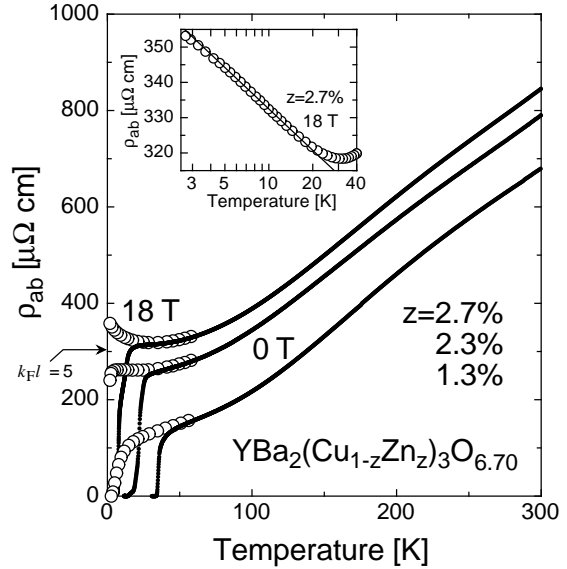}$$
\end{center}
\caption{$\rho_{\rm ab}(T)$ of $y$=6.70 crystals
with different $z$ (2.7\%, 2.3\%, and 1.3\%)
in 0 T (solid lines) and in 18 T (open circles).
Inset: $\log T$ plot of $\rho_{\rm ab}(T)$ for the $z$=2.7\%
sample to show the $\log(1/T)$ dependence (straight line).
The deviation at the lowest temperatures is due to superconducting
fluctuations.}
\label{fig1}
\end{figure}

Figure 2 (a) and (b) show the temperature dependence of $\rho_{\rm ab}$
in various fields
for the $y$=6.70, $z$=2.7\% sample and the $y$=6.70, $z$=2.3\% sample,
respectively.
The magnetic-field dependence of $\rho_{ab}$ at a representative
temperature is shown in the insets.
Apparently, 18-T field is sufficient to suppress superconductivity
above 5 K in the $z$=2.7\% sample and above 10 K in the $z$=2.3\% sample.
One may notice in Fig. 2 (b) that the 18-T data show a very slight
upturn below 15 K, which suggests that this $z$=2.3\% sample might show
stronger upturn in much higher fields.

By analogy with the metal-insulator crossover observed in the
low-temperature normal-state of LSCO \cite{Boebinger},
one may expect that increasing oxygen content makes the $z$=2.7\%
sample metallic.  This is actually the case, as is shown in Fig. 3,
where $\rho_{\rm ab}(T)$ of $z$=2.7\% samples with $y$=6.70, 6.75, 6.80,
and 6.83 are plotted.
When the oxygen content exceeds 6.80, the low-temperature
upturn in $\rho_{\rm ab}(T)$
disappears and a metallic behavior is observed in the $y$=6.83 sample
in 15 T.
It should be noted that the localization behavior is observed here
in samples with quite small resistivity.
For example, the minimum of $\rho_{ab}(T)$ for the $y$=6.80, $z$=2.7\%
sample is about 180 $\mu\Omega$cm.  This value corresponds to
$k_{F}l\simeq$8.5, if we use the formula $l=hc_0/\rho k_{F}e^2$ for 2D
electrons \cite{Boebinger} ($c_0$ is the $c$-axis unit length);
therefore, the localization behavior is taking place in quite a
clean system where localization is not normally observed
\cite{Mott}.
In this sense, the localization behavior observed here is quite
unusual.

\vspace{-0.7cm}
\begin{figure}[htbp]
\begin{center}
 \epsfxsize=75mm
 $$\epsffile{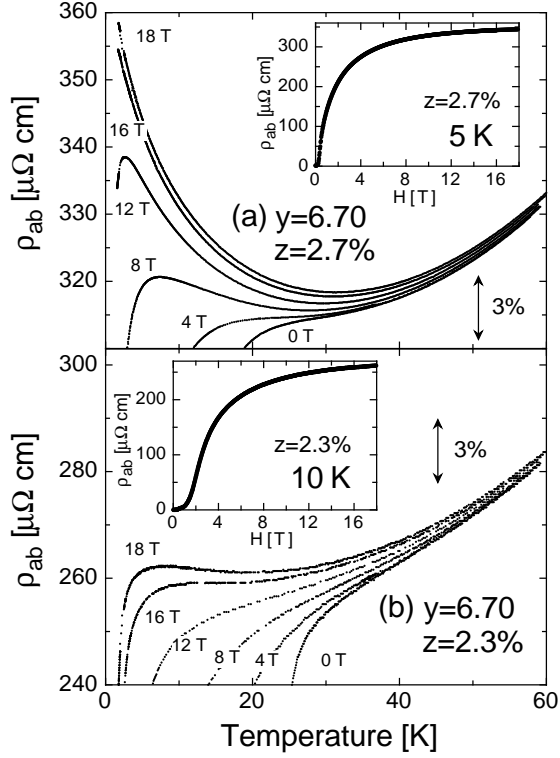}$$
\end{center}
\caption{$\rho_{\rm ab}(T)$ in 0, 4, 8, 12, 16 and
18 T for (a) $y$=6.70, $z$=2.7\% sample and
(b) $y$=6.70, $z$=2.3\% sample. Insets show the
$H$ dependence of $\rho_{\rm ab}$.}
\label{fig2}
\end{figure}

\vspace{-0.7cm}
\begin{figure}[htbp]
\begin{center}
 \epsfxsize=75mm
 $$\epsffile{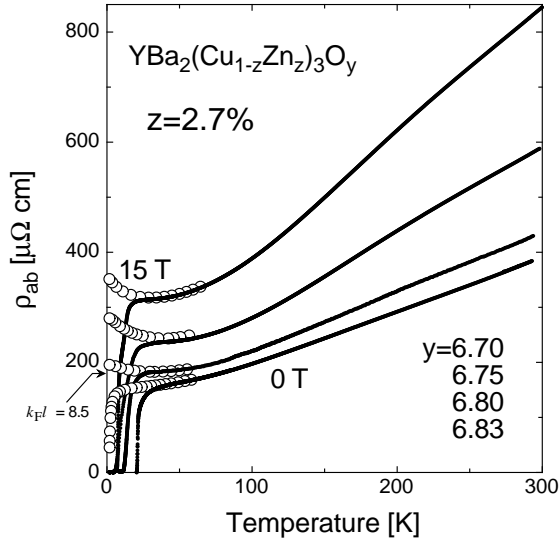}$$
\end{center}
\caption{$\rho_{\rm ab}(T)$ in 0 and 15 T
for $z$=2.7\% samples with y=6.70, 6.75, 6.80, and 6.83.}
\label{fig3}
\end{figure}

The inset to Fig. 4 shows a comparison of $\rho_{\rm ab}(T)$ of two
samples, $y$=6.70, $z$=2.3\% (sample M1) and
$y$=6.80, $z$=2.7\% (sample I1), to demonstrate that the onset of the localization
behavior is not correlated with residual resistivity
$\rho_{\rm res}$.
Sample M1 has apparently higher $\rho_{\rm res}$ than that of sample I1,
indicating there is no critical $\rho_{\rm res}$ for the localization
behavior to take place.
This observation is in contrast to the zero-field S-I transition
observed in YBCO and LSCO \cite{Fukuzumi}, which occurs whenever the
resistivity exceeds the critical value, as mentioned earlier.
It should be noted that all the samples which show
the localization behavior in
Figs. 1 and 3 have smaller resistivity than the critical value
400 $\mu \Omega$cm \cite{Fukuzumi}
and thus are on the superconductor side of the
S-I transition; in fact, all those samples are superconducting
and do not show any
upturn in $\rho_{ab}$ in zero field, which is
in accord with the zero-field S-I transition picture.
What is new here is that the unusual localization behavior shows up
once the superconductivity is suppressed, where charge is no more
carried by Cooper pairs.
The result shown in the inset to Fig. 4 indicates that the onset of the
unusual localization behavior is determined primarily by the
Zn concentration but not by $\rho_{\rm res}$.

\vspace{-0.7cm}
\begin{figure}[htbp]
\begin{center}
 \epsfxsize=75mm
 $$\epsffile{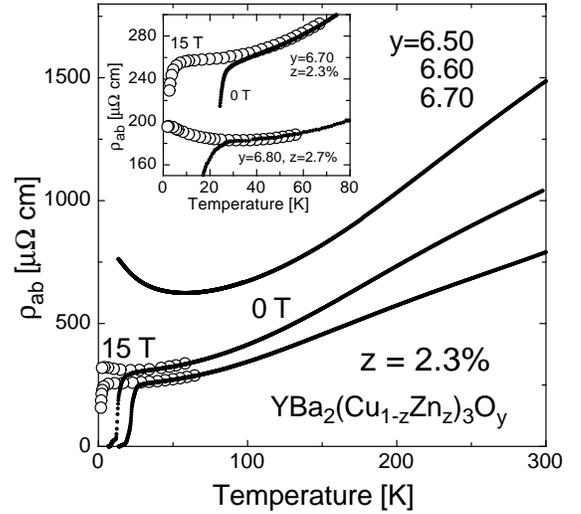}$$
\end{center}
\caption{$\rho_{\rm ab}(T)$ in 0 and 15 T
for $z$=2.3\% samples with y=6.50, 6.60, and 6.70.
Inset: $\rho_{ab}(T)$ in 0 and 15 T for
$y$=6.70, $z$=2.3\% sample and
$y$=6.80, $z$=2.7\% sample.}
\label{fig4}
\end{figure}

The evolution of $\rho_{ab}(T)$ with changing oxygen content for
$z$=2.3\% is shown in the main panel of Fig. 4.
For this Zn concentration,
the charge transport has much weaker tendency towards localization;
even $y$=6.60 sample shows only a slight upturn in 15 T.
When the oxygen content is further reduced to 6.50,
$\rho_{\rm res}$ exceeds the critical value for the S-I transition
and $\rho_{ab}(T)$ shows an insulating behavior already
in zero field.
Comparison between the $z$=2.7\% samples (Fig. 3) and the
$z$=2.3\% samples (Fig. 4) lead to the conclusion that
the tendency toward carrier localization in the high-field normal state
becomes much stronger when the Zn concentration is increased from
2.3\% to 2.7\%.
Note that
there is no drastic difference between the $z$=2.7\% samples and the
$z$=2.3\% samples if we only look at the zero-field data.

Now let us discuss the possible origin of the unusual localization
behavior.
What we observed in Zn-doped YBCO has strong similarities to
the unusual localization behavior found in underdoped LSCO
\cite{Ando,Boebinger,MOS}:
In both systems, the localization behavior
takes place in a region of \lq\lq clean" charge transport where
$k_{F}l$ is larger than 5;
the temperature dependence of the resistivity
is consistent with $\log(1/T)$ and the size of the increase in resistivity
is larger than what is expected from weak localization;
the localization behavior shows up at low temperatures
only when the superconductivity is suppressed with high magnetic fields.
Therefore, it is naturally expected that the localization behaviors
observed in the two systems have a common origin.
When we look for some common features in the underlying
electronic/magnetic structures of the two systems,
we hit on the fact that, as mentioned earlier, both systems have
some kind of \lq\lq defects" to the AF correlation in the CuO$_2$ planes;
LSCO system is likely to have dynamical charge
stripes which separate AF domains, while the Zn impurity in YBCO system
acts like a point-like defect in the AF correlation.
Note that both the dynamical charge stripes and the Zn-substitution
effect are most strongly seen at excitations with wave vectors near
${\bf q}$=$(\pi,\pi)$.

The above considerations lead to the conclusion that the local
destruction of the AF correlation in the CuO$_2$ planes might be
the origin of the unusual insulating behavior.
This conclusion in turn suggests that the dynamical AF correlation
in the CuO$_2$ planes governs the peculiar normal-state charge
transport of the cuprates.
If so, the charge in cuprates may be carried by a quasiparticle which
is closely related to the AF correlation, and a local disturbance of
the AF correlation, in the presence of a high magnetic field,
may strongly scatter the quasiparticle and brings about the unusual
charge localization.
The fact that the localization behavior is much stronger in the LSCO
system (where there is more than a factor of 2 increase in resistivity
at low temperatures in 60 T \cite{Ando})
may mean that the scattering from 1D defects is much stronger
than that from point-like defects.

In summary, we measured the in-plane resistivity of Zn-doped
YBCO crystals down to low temperatures by suppressing superconductivity
with high magnetic fields.
It is found that the Zn substitution induces unusual carrier
localization at low temperatures in a \lq\lq clean"
system with $k_{F}l$ as large as 8.5, and
the tendency toward carrier localization becomes much stronger
when the Zn concentration is increased from 2.3\% to 2.7\%.
Strong similarities are found between the localization behavior observed
here and the
$\log(1/T)$ behavior observed in underdoped LSCO.
Examination of the common characteristics in LSCO and Zn-doped YBCO
leads to the conclusion that the local destruction of the
AF correlation either by 1D defects (dynamical charge stripes in LSCO)
or by point-like defects (Zn in YBCO) might be the origin of the
unusual charge localization.

%



\begin{references}
\bibitem{Maple}
For a recent review, see M.B. Maple, cond-mat/9802202
(unpublished).
\bibitem{Yamada}
K. Yamada, C.H. Lee, K. Karahashi, J. Wada, S. Wakimoto, S. Ueki,
H. Kimura, Y. Endoh, S. Hosoya, G. Shirane, R.J. Birgeneau, M. Greven,
M.A. Kastner, and Y.J. Kim, Phys. Rev. B 57, 6165 (1998).
\bibitem{Mook}
P. Dai, H.A. Mook, and F. Dogan, Phys. Rev. Lett. 80, 1738 (1998).
\bibitem{Mahajan}
A.V. Mahajan, H. Alloul, G. Collin, and J.F. Marucco, Phys. Rev. Lett.
72, 3100 (1994), and references therein.
\bibitem{Chien}
T.R. Chien, Z.Z. Wang, and N.P. Ong, Phys. Rev. Lett. 67, 2088 (1991).
\bibitem{Fukuzumi}
Y. Fukuzumi, K. Mizuhashi, K. Takenaka, and S. Uchida,
Phys. Rev. Lett. 76, 684 (1996).
\bibitem{Mizuhashi}
K. Mizuhashi, K. Takenaka, Y. Fukuzumi, and S. Uchida,
Phys. Rev. B 52, R3884 (1995).
\bibitem{Ando}
Y. Ando, G.S. Boebinger, A. Passner, T. Kimura, and K. Kishio,
Phys. Rev. Lett. 75, 4662 (1995).
\bibitem{Tranquada}
J.M. Tranquada, P. Wochner, and D.J. Buttrey,
Phys. Rev. Lett. 79, 2133 (1997).
\bibitem{Suzuki}
T. Suzuki, T. Goto, K. Chiba, T. Shinoda, T. Fukase, H. Kimura,
K. Yamada, M. Ohashi, and Y. Yamaguchi, Phys. Rev. B 57, R3229 (1998).
\bibitem{Tranquada2}
J.M. Tranquada, J.D. Axe, N. Ichikawa, Y. Nakamura, S. Uchida,
and B. Nachumi, Phys. Rev. B 54, 7489 (1996); C. Castellani,
C. Di Castro, and M. Grilli, cond-mat/9709278 (unpublished).
\bibitem{Alloul}
H. Alloul, P. Mendels, H. Casalta, J.F. Marucco, and J. Arabski,
Phys. Rev. Lett. 67, 3140 (1991).
\bibitem{Walker}
D.J.C. Walker, A.P. Mackenzie, and J.R. Cooper,
Phys. Rev. B 51, 15653 (1995).
\bibitem{Fisher}
M.P.A. Fisher, G. Grinstein, and S.M. Girvin, Phys. Rev. Lett.
64, 587 (1990).
\bibitem{Nagaosa}
N. Nagaosa and P.A. Lee, Phys. Rev. Lett. 79, 3755 (1997).
\bibitem{Boebinger}
G.S. Boebinger, Y. Ando, A. Passner, K. Tamasaku, N. Ichikawa, S. Uchida,
M. Okuya, T. Kimura, J. Shimoyama, and K. Kishio,
Phys. Rev. Lett. 77, 5417 (1996).
\bibitem{Mott}
N. F. Mott, {\it Metal-Insulator Transitions, 2nd edition}
(Taylor \& Francis, London, 1990).
\bibitem{MOS}
Y. Ando, G.S. Boebinger, A. Passner, K. Tamasaku, N. Ichikawa, S. Uchida,
M. Okuya, T. Kimura, J. Shimoyama, and K. Kishio,
J. Low Temp. Phys. {\bf 105}, 867 (1996).

\end{references}
\end{document}